\documentstyle[revtex,preprint,eqsecnum]{aps}
\newcommand{\F}{{\cal F} }
\newcommand{\K}{{\cal K} }
\newcommand{\M}{{\cal M} }
\renewcommand{\O}{{\cal O} }
\renewcommand{\S}{{\cal S} }

\newcommand{\ben}{\begin{equation}}
\newcommand{\een}{\end{equation}}
\newcommand{\bea}{\begin{eqnarray}}
\newcommand{\eea}{\end{eqnarray}}

\newcommand{\e}{\epsilon}
\newcommand{\parent}{p}
\newcommand{\hard}{h}
\newcommand{\unobserved}{u}
\newcommand{\MSbar}{\overline{\rm MS}}
\newcommand{\smin}{s_{\rm min}}
\newcommand{\dPhase}{d\,\!P}
\newcommand{\LO}{{\rm LO}}
\newcommand{\NLO}{{\rm NLO}}
\newcommand{\deltasigma}{\delta\sigma}
\tightenlines
\begin{document}

\pagestyle{myheadings}
\markboth{December 3, 1992}{December 3, 1992}
\typeout{}

\preprint{FERMILAB-Pub-92/230-T\vspace{3mm}}
\preprint{DTP/92/64\vspace{3mm}}
\preprint{CERN-TH 6750/92\vspace{3mm}}
\preprint{hep--ph/9302225\vspace{3mm}}

\vspace{1cm}
\begin{title}
Higher Order Corrections to Jet Cross Sections in Hadron Colliders
\end{title}
\author{
W.~T.~Giele\cite{SSCfell}}
\begin{instit}
Fermi National Accelerator Laboratory, P.O.~Box 500,\\
Batavia, IL 60510, U.S.A. \\
{\tt giele@fnth02.fnal.gov}\\
\end{instit}
\author{E.~W.~N.~ Glover}
\begin{instit}
Physics Department, University of Durham, \\
Durham DH1~3LE, England\\
{\tt ewng@hep.dur.ac.uk}\\
\end{instit}
\author{David~A.~Kosower}
\begin{instit}
Theory Division, CERN,\\
CH-1211 Gen\`eve 23, Switzerland\\
{\rm and}\\
Service de Physique Th\'eorique\cite{saclay}, Centre d'Etudes de Saclay,\\
F-91191 Gif-sur-Yvette cedex, France\\
{\tt kosower@amoco.saclay.cea.fr}\\
\end{instit}
\begin{abstract}
We describe a general method of calculating the fully differential cross
section for the production of jets at next-to-leading order in a hadron
collider.
This method is based on a `crossing' of next-to-leading order calculations
with all partons in the final state.
The method introduces universal crossing functions that
allow a modular approach to next-to-leading order calculations for any process
with initial state partons.
These techniques are applied to the production of jets in association
with a vector boson including all decay correlations of the final
state observables.
\end{abstract}

\section{Introduction}
\label{sec:intro}

One of the most striking features of hadronic events is the appearance
of `jets' of hadrons.  By use of a suitable experimental jet
algorithm, the hadronic data may be organized into final states
containing a definite number of jets.  This defines the topological
structure of the event for a given jet algorithm.  Different jet
algorithms or jet defining cuts can lead to a different number of jets
being observed in a given event.  As a result, jet cross sections
depend on the procedure used to define an experimental jet.
Nevertheless, at the experimental level, a jet is a perfectly well
defined quantity since for a given jet algorithm each event contains a
precisely determined number of jets.

{}From the theoretical point of view, the jet algorithm plays an
important role in selecting high momentum transfer events in which the
soft radiation is removed by a cut on the minimal transverse energy of
the jet.  At the same time, individual hadron behavior is averaged out
by replacing all hadrons within a cone of a given size by a single jet
axis and jet energy. Because the hadronic information is averaged out,
we can relate the hadronic jet axis and energy observed in the
experiment to a jet axis and energy constructed from a parton shower
calculated within perturbative QCD .  The jet axis and energy obtained
{}from the hadronic shower are thus modelled by the jet axis and energy
obtained from the parton shower. This is a weak form of the
parton-hadron duality theorem \cite{lphd}. Of course, non-perturbative
hadronization effects are not predicted by perturbative QCD.
Similarly, the contributions from the underlying event (at hadron
colliders) are not included. `Sensible' jet algorithms minimize these
effects and allow a more direct comparison between theory and
experiment.

More precise theoretical predictions of jet cross sections are
potentially valuable since new physics is often evident in
events containing a specific number of jets.  The corresponding QCD
background is then the exclusive jet cross section containing the same
number of jets.  For example, the signal to background ratio for the
top quark in the lepton + jets channel is improved by demanding that
more jets be observed
\cite{top}.

The lowest order matrix elements for the two most prominent processes
containing jets at hadron colliders,
\bea
p\bar p &\to & n_1 ~{\rm jets},\\
p\bar p &\to & W^\pm/Z + n_2 ~{\rm jets},
\eea
have been computed for $n_1
\leq 5$ \cite{5j} and $n_2 \leq 4$ \cite{vc4} by making use of helicity
amplitudes \cite{hel}, color
decompositions \cite{six,MPX,cod} and recursion relations
\cite{rec1,BKT} to control the rapid increase in the number of Feynman
diagrams as the number of partons involved grows.  The jet cross
section is then obtained by Monte Carlo integration of the phase space
of the final state partons.  This approach allows any experimental jet
algorithm and acceptance to be applied, and one can study any
distribution depending on the jet observables.  It is important to
note that at leading order, the jet is modelled by a single parton.
The jet defining cuts are applied to this lone parton and the parton's
direction and energy describe the jet's axis and energy \cite{foot1}.

Comparisons of lowest order QCD predictions of jet distributions with the data
have proved reasonable, bearing in mind the fact that one is comparing a
theoretical perturbative calculation with hadronic data.
Generally, the shapes of infrared safe distributions are well
predicted while the overall normalization is uncertain, due to a dependence on
the unphysical
renormalization and factorization scales $\mu_R$ and $\mu_F$ \cite{expscale}.

The addition of next-to-leading order effects produces three important
improvements over a leading order calculation. First, the dependence
on the unphysical scales $\mu_R$ and $\mu_F$ is reduced so that the
normalization is more certain.  Second, we begin to reconstruct the
parton shower.  This means that two partons may combine to form a
single jet.  As a result, jet cross sections become sensitive to the
details of the jet finding algorithm, particularly the way in which
the hadrons are combined to form the jet axis and energy, and to the
size of a jet cone.  This sensitivity is also seen in experimental
results.  Third, the calculation becomes more sensitive to detector
limitations, because radiation outside the detector is simulated. This
can change leading order results considerably for quantities such as
the missing transverse energy in events containing a $W$ boson.

\newcommand{\yIR}{y_{\rm IR}}
{}From a more theoretical point of view, the presence of infrared
logarithms in a generic perturbative QCD prediction implies that the
expansion of physical quantities is not strictly a series in the
coupling constant $\alpha_s$, but is rather a series in $\alpha_s
\ln^2 \yIR$ and $\alpha_s \ln \yIR$ (as well as $\alpha_s$ alone),
where $\yIR$ is an experimental resolution (for example, the minimum
jet invariant mass in $e^+e^-$ collisions).  Thus even in the
perturbative regime, the leading-order result --- where the size of
these logarithms is uncalculated --- suffers from potentially large
corrections which might spoil the applicability of perturbative QCD.
In a next-to-leading order calculation, these logarithms are
calculated explicitly, and thus one regains confidence in the
applicability of the perturbative expansion.  From this purist's point
of view, a next-to-leading order calculation is necessary in order to
understand whether a leading-order result is trustworthy.

At next-to-leading order, the $n$-jet cross section receives
contributions from virtual corrections to $n$-parton, and from real
corrections in the form of $(n+1)$-parton, final states.  Both
contributions are divergent.  The matrix elements for the virtual
diagrams are infrared-divergent, while the real $(n+1)$-parton matrix
elements are well defined.  However, when the $(n+1)$-parton matrix
elements are integrated over the allowed regions of phase space an
infrared-divergent cross section is obtained.  This comes about
because the jet algorithm allows one of the partons to be soft, or for
two partons within a jet cone to be collinear.

In order to cancel these divergences explicitly, it is convenient to
divide the $(n+1)$-parton phase space into regions where
$(n+1)$-partons are `resolved' and regions where only $n$-partons are
`resolved' \cite{owens,greco,GG,foot2}.  For example, if the invariant
mass of two partons, $s_{ij}$, is smaller than some theoretical parton
resolution parameter $\smin$ only one parton is resolved, while if
$s_{ij} > \smin$ both partons are resolved.  All of the divergences
{}from the $(n+1)$-parton final state separate and are associated with
the regions where only $n$-partons are resolved.  These divergences
can be cancelled directly against the virtual corrections to the
$n$-parton cross section. With such a physical picture \cite{foot3} it
is straightforward to extend this method to deal with any number of
partons in the final state.  Indeed, using a color decomposition of
the amplitude, one can write down a simple soft factorization
\cite{soft} for the sub-amplitudes, which in turn allows the
construction of a universal set of functions summarizing the soft and
collinear behavior of the matrix element of any colored particles,
both before and after the cancellation of infrared divergences
described above. Recently, we have described how this scheme may be
applied to multijet cross sections in $e^+e^-$ annihilation \cite{GG}.
This case is rather special since there can be no QCD radiation in the
initial state and all divergences therefore reside in the final state.
In the paper, we wish to extend this method to processes which involve
partons in the initial state.

In Sec.~\ref{sec:org}, we show how to set up the next-to-leading order
calculation of the matrix elements for $e^+e^- \to n$~partons and how
they may be `crossed' to give the cross section for $p\bar p \to V +
(n-2)$~partons where $V = W^\pm,~\gamma^*,~Z$.  In order to do this it
is convenient to introduce (in addition to the universal soft and
collinear functions mentioned above) a set of universal crossing
functions which multiply the lowest order cross section.  These
crossing functions are essentially convolutions of the input structure
functions with the Altarelli-Parisi splitting functions.
Section~\ref{sec:init} deals with the derivation of the crossing
functions which enable us to cross final state partons to the initial
state.  In Sec.~\ref{sec:app}, we construct explicit Monte Carlo
programs for,
\begin{equation}
p\bar p \to V + 0,~1~{\rm jets} \to \ell \bar \ell' + 0,~1~{\rm jets} ,
\end{equation}
at next-to-leading order. The jet algorithm may then be applied
directly to the $n$ and $(n+1)$ parton final states.  All dependence
on the unphysical parton resolution parameter $\smin$
cancels numerically .  The cross section is fully differential in all
jet and lepton observables and therefore differs from calculations of
the $W$ transverse momentum distribution at ${\cal O}(\alpha_s^2)$
\cite{AR,Gon} or of the single jet inclusive transverse momentum
distribution at ${\cal O}(\alpha_s^3)$ \cite{greco,EKS}.
Equivalent techniques have been applied to
$p\bar p \to W^\pm + 0$~jets \cite{BR} and $p\bar p \to 2$~jets
\cite{EKS2} at next-to-leading order.
Finally, we summarize our results in Sec.~\ref{sec:con}.

\section{Calculational organization}
\label{sec:org}

In this section we give a general overview of the manner in which the
calculation is set up, without going into details, which we shall
present in subsequent sections. The organization minimizes the
computational effort while retaining the standard $\MSbar$
prescription \cite{MSbar}.  For example, the cumbersome
$d$-dimensional squaring of the matrix elements is avoided.
Furthermore, the introduction of universal crossing functions will
allow us to obtain the initial state parton cross sections from the
all-outgoing cross section.  These techniques depend crucially on the
universality of the QCD soft and collinear radiation patterns.

An efficient way of organizing next-to-leading order calculations for
all-outgoing parton processes was given in ref. \cite{GG}. The basis of
this method is the use of ordered amplitudes associated with each
color structure rather than the full matrix element. For example, let
us consider the decay of a vector boson into a quark-antiquark pair
with $n$ accompanying gluons. The full squared amplitude is obtained
by summing the squared ordered amplitudes over all permutations of the
gluons \cite{six,MPX,cod},
\begin{equation}
\left|{\cal M}_V(V\rightarrow q\bar{q}\,+\,n\ g)\right|^2\sim
\sum_{\it Perm.} \left| A(q;g_1,\ldots,g_n;\bar{q})\right|^2\ .
\label{eq:m2v}
\end{equation}
For simplicity, we keep only the terms at leading order in the number
of colors.  See Sec.~\ref{sec:app} for a full discussion of the
subleading terms.  Of crucial importance is the fact that the ordered
amplitudes exhibit QED-like factorization \cite{YFS} in the soft and
collinear limits \cite{soft}.  This forms the basis of our method and
allows the integration over the singular (or unresolved) parts of
phase space without calculating the hard matrix element explicitly,
\begin{equation}
\int d\, P_{\rm soft/collinear}
\left| A(q^\prime;g_1^\prime,\ldots,g_{n+1}^\prime;\bar{q}^\prime)
\right|^2\rightarrow R(q;g_1,\ldots,g_n;\bar{q}|\smin)
\left| A(q;g_1,\ldots,g_{n};\bar{q}) \right|^2
\label{eq:easyr}
\end{equation}
where the soft and collinear regions of phase space are defined by the
invariant mass cut $\smin$. At this point everything is done in an
arbitrary number of dimensions \cite{HV,DR}.  However, we never have to
calculate the hard amplitude squared explicitly since this result is
obtained without any detailed knowledge of the hard process.  The next
step is to calculate the virtual corrections to the squared matrix
elements which have the generic form,
\begin{equation}
\left|{\cal M}_V(V\rightarrow q\bar{q}\,+\,n\ g)\right|^2_V =  \sum_{Perm.}
V(q;g_1,\ldots,g_n;\bar{q})\ \left| A(q;g_1,\ldots,g_n;\bar{q})\right|^2 +
{\cal F}(q;g_1,\ldots,g_n;\bar{q}),
\end{equation}
where $V(q;g_1,\ldots,g_n;\bar{q})$ is the singular part proportional
to the tree level ordered amplitude and ${\cal
F}(q;g_1,\ldots,g_n;\bar{q})$ is the remaining finite contribution.
This can be immediately combined with the unresolved phase space
contribution Eq.~\ref{eq:easyr} to give the finite next-to-leading
order squared matrix elements,
\begin{equation}
\left| {\cal M}_V(V\rightarrow q\bar{q}\,+\,n\ g)\right|^2_F
\sim\sum_{Perm.}\left(\left[1 + \K\right]
\left|A(q;g_1,\ldots,g_n;\bar{q})\right|^2+
{\cal F}(q;g_1,\ldots,g_n;\bar{q})\right).
\label{eq:final}
\end{equation}
Note that due to the Kinoshita-Lee-Nauenberg \cite{KLN,GS} theorem the
combination of the phase space factor and the virtual factor, $\K = R
+ V$, is finite.  As a direct consequence we can now perform the
squaring and summation over the polarizations in 4 dimensions using
the standard techniques developed for evaluating complicated tree
level amplitudes (helicities \cite{hel} and recurrence relations
\cite{rec1}).  While $V$ can be calculated in a process-independent
manner \cite{GG}, the finite remainder of the virtual correction ${\cal F}$
needs to be calculated on a process-by-process basis.  The general
structure is process-independent, and it is in this sense that the
$\K$ factor is universal.

In order to generalize the framework above, used in $e^+ e^-$
collisions, to hadronic collisions, we must include initial state
partons in the calculation.  One useful property of lowest
order matrix elements is that of ``crossing''.  In other words, the
matrix elements for $V\to q\bar{q}\ +\ n\,g$ are related to those for
the crossed processes,
\begin{eqnarray}
q\bar{q} &\to& V\ +\ n\,g, \nonumber \\ qg &\to& V\ +\ q\ +\ (n-1)\,g,
\\
\bar{q}g &\to& V\ +\ \bar{q}\ +\ (n-1)\,g, \nonumber \\
gg &\to& V\ +\ q\bar{q}\ +\ (n-2)\,g, \nonumber
\end{eqnarray}
by reversal of the momentum and helicity of the crossed particles.  The
fully differential cross section at leading order in the collision of
hadrons $H_1$ and $H_2$,
\begin{equation}
H_1\ +\ H_2\to V\ +\ n\,\mbox{partons},
\end{equation}
is,
\begin{equation}
d\,\sigma_{H_1H_2} = \sum_{ab} f_a^{H_1}(x_1)f_b^{H_2}(x_2)
d\,\sigma^{\LO}_{ab}(x_1,x_2)\ d\,x_1 d\,x_2.
\label{eq:siglo}
\end{equation}
Here $f_a^{H}(x)$ is the probability density of finding parton $a$ in
hadron $H$ with momentum fraction $x$ and,
\begin{equation}
d\,\sigma^{\LO}_{ab}(x_1,x_2) = \frac{\Phi_{ab}}{2\,s_{ab}} \left|{\cal
M}_{ab}\right|^2\ d\,P(ab\to V\ +\ n\,\mbox{partons}),
\end{equation}
where $\Phi_{ab}$ is the appropriate spin and color averaging factor
and $d\,P$ the $V\ +\ n$ parton phase space where all parton pairs
satisfy $s_{ij} > \smin$. The matrix elements for $ab\to V + n$
partons are denoted $\left|{\cal M}_{ab}\right|^2$ and are related
by crossing to $\left|{\cal M}_{V}\right|^2$, Eq.~\ref{eq:m2v}.

We have already discussed how the next-to-leading order matrix
elements for $e^+e^- \to n$~partons can be written in an explicitly
finite way using the parton resolution parameter $\smin$.  We now wish
to extend this to processes involving partons in the initial state
while maintaining the crossing properties of lowest order.  In order
to achieve this, the next-to-leading order hadronic cross section must
be defined by,
\begin{equation}
d\,\sigma_{H_1H_2} = \sum_{ab} \F_a^{H_1}(x_1)\F_b^{H_2}(x_2)
d\,\sigma^{\NLO}_{ab}(x_1,x_2)\ d\,x_1 d\,x_2,
\label{eq:signlo}
\end{equation}
where $\F_a^H$ is the ``effective'' next-to-leading order structure
function and $d\sigma^{\NLO}_{ab}$ the ``crossed'' analogue of the
finite next-to-leading order partonic cross section.  This cross
section can be expanded as a series in the coupling constant,
\begin{equation}
d\,\sigma^{\NLO}_{ab} = d\,\sigma^{\LO}_{ab}+ \alpha_s
d\,\deltasigma^{\NLO}_{ab} + \O (\alpha_s^2),
\end{equation}
where we have extracted the coupling constant from the finite crossed
matrix elements.  Note that $\alpha_s$ is evaluated at the
renormalization scale $\mu_R$.  Similarly, after mass factorization,
the effective structure function $\F^H_a$ may be written as,
\begin{equation}
\F^H_a(x) = f^H_a(x,\mu_F) + \alpha_s\,C^H_a(x,\mu_F)
+ \O (\alpha_s^2),
\label{eq:F}
\end{equation}
where $\mu_F$ is the factorization scale.  Both $f^H_a(x,\mu_F)$ and
the crossing function $C^H_a(x,\mu_F)$ are finite.
Once again $\alpha_s$ is evaluated at the renormalization scale. In principle
one could evaluate $\alpha_s$ at the mass factorization scale, however,
provided $\alpha_s\log(\mu_R^2/\mu_F^2) \ll 1$,
the difference is of $\O(\alpha_s^2)$ and can be ignored.
For a detailed derivation of the structure of the
crossing function $C^H_a$ we refer the reader to Sec.~\ref{sec:init}.

Inserting these definitions back into Eq.~\ref{eq:signlo} and
expanding up to $\O(\alpha_s)$ we find,
\begin{eqnarray}
d\,\sigma_{H_1H_2} & = & \sum_{ab} \Biggl[f_a^{H_1}(x_1)f_b^{H_2}(x_2)
\left\{ d\,\sigma^{\LO}_{ab}(x_1,x_2)+ \alpha_s\,
d\,\deltasigma^{\NLO}_{ab}(x_1,x_2)
\right\}\nonumber \\
&+& \alpha_s\left\{C^{H_1}_a(x_1)f^{H_2}_b(x_2)
+f^{H_1}_a(x_1)C^{H_2}_b(x_2)\right\} d\,\sigma^{\LO}_{ab}(x_1,x_2) +
\O (\alpha_s^2)\Biggr]
\ d\,x_1 d\,x_2 .
\label{eq:nlo} \nonumber \\
\end{eqnarray}
For simplicity, we have suppressed the dependence on the renormalization scale
in the coupling constant and the factorization
scale in the structure and crossing functions.

The crossing function receives two contributions which both stem from
the fact that we consider two partons to be unresolved when their
invariant mass is smaller than the parton resolution parameter
$\smin$.  Firstly, we cannot distinguish between a single initial
state parton and a parton which emits collinear radiation such that
the invariant mass of the collinear pair is smaller than $\smin$.
This implies that part of the initial state collinear radiation is
removed from the hard scattering and absorbed into the effective
structure function.  Clearly, this contribution depends on $\smin$ and
therefore so do the crossing functions (and also $\F^H_a$).  This term
is a convolution of $f^H_a$ with the Altarelli-Parisi splitting
function, $P_{a\to c}(z)$ \cite{AP}.

The second contribution arises from crossing a pair of collinear
partons with an invariant mass smaller than $\smin$ from the final
state to the initial state.  In principle we should remove this
contribution from $d\,\deltasigma^{\NLO}_{ab}$, however in order to
preserve the structure of Eq.~\ref{eq:signlo}, we subtract this
contribution from the parton density function.  This is possible
because we cannot distinguish the two parton incoming state with
invariant mass smaller than $\smin$ from a single incoming parton.

Both of these contributions are divergent and schematically,
\begin{equation}
C^H_a(x) \sim \sum_c \left [ \int_x^1 \frac{dz}{z} f^H_{c}
\left(\frac{x}{z}\right)P_{c\to a}(z) - f^H_{a}(x) \int_0^1 dz\,
P_{a\to c} (z) \right]
\frac{\smin^{-\e}}{\e}.
\label{eq:C}
\end{equation}
A more precise formulation of the crossing function is given in the
next section including all $d$-dimensional factors.  After mass
factorization the crossing function is rendered finite,
\begin{equation}
C^{H,\rm scheme}_a (x,\mu_F) = \left(\frac{N}{2\pi}\right)
\left [ A^H_a (x) \log\left(\frac{\smin}{\mu_F^2}\right)
      + B_a^{H,\rm scheme}(x)
 \right].
\label{eq:Cab}
\end{equation}
Although $A^H_a$ is scheme independent, $B^H_a$ does depend on the
mass factorization scheme and therefore so does $C^H_a$.  Explicit
forms for these functions in the $\MSbar$ scheme are given in
the next section.

The overall cross section cannot depend on the unphysical parameter
$\smin$.  When the contribution from $H_1 + H_2 \to V + (n+1)$~partons
where all partons are resolved is included, the $\smin$ in the
logarithm is replaced by an energy scale defined by the experimental
cuts.  In a numerical computation, one would also force the
factorization scale to be determined by the experimental cuts; the
argument of the logarithm would then be of $\O(1)$ and the contribution
{}from $A^H_a$ would be small.  If this were not true, the
logarithm would be large so that $\alpha_s A^H_a(x)
\log\left( E_{\rm exp}^2/\mu_F^2\right) \simeq f^H_a(x)$,
perturbation theory would break down and a resummation of the leading
logarithms would be necessary.

\section{Derivation of the crossing functions}
\label{sec:init}

In this section we derive explicit formul\ae\ for the crossing
functions $C_a^H(x,\mu_F)$ as defined in
eqs.~\ref{eq:signlo}-\ref{eq:nlo}. First, we derive the initial state
collinear phase space behavior in a parametrization suitable for our
parton resolution parameter. We then reformulate the standard
collinear matrix element factorization in the ordered amplitude
language. Combining these results enables us to derive the universal
crossing functions which after mass factorization yield finite
crossing functions.

\subsection{The initial-state collinear behavior of phase space}

First consider the production of a heavy object, $Q$, by the collision
of two massless particles with momenta $p_a$ and $p_\hard$.  The
$d$-dimensional phase space measure, including the flux factor, is
given by,
\begin{equation}
\frac{1}{2s_{a\hard}}dP^d(a+\hard \to Q) =
\frac{2\pi}{2s_{a\hard}}\delta(s_{a\hard}-Q^2),
\end{equation}
where $s_{a\hard}= (p_a+p_\hard)^2$.  This extends straightforwardly
to the production of any number of particles (massless or otherwise)
with momenta $p_1,\ldots,p_n$ by use of the relation,
\begin{equation}
dP^d(a+\hard \to 1+ \cdots +n) = dP^d(a+\hard \to Q) \frac{dQ^2}{2\pi}
dP^d(Q\to 1+\cdots + n).
\end{equation}

Next consider the phase space for the production of a massless
particle with momentum $p_\unobserved$ in association with $Q$ from
the collision of two massless particles with momenta $p_a$ and
$p_\parent$,
\begin{equation}
\frac{1}{2s_{a\parent}}dP^d(a+\parent \to \unobserved + Q) = (2\pi)^{2-d}
\frac{d|s_{a\unobserved}|\,d|s_{\parent\unobserved}|}{s_{a\parent}^2}
\left[\frac{|s_{a\unobserved}||s_{\parent\unobserved}|}
{s_{a\parent}}\right]^{-(2-d/2)}
\frac{d\Omega_{d-3}}{8}\ \delta(s_{a\parent}-|s_{a\unobserved}|-
|s_{\parent\unobserved}|-Q^2),
\end{equation}
where we integrate over the invariant mass of $Q$ and the polar angle
with respect to $p_\parent$ by using $|s_{a\unobserved}|$ and
$|s_{\parent\unobserved}|$, as well as integrating over the $(d-3)$
azimuthal angles relative to the direction of $p_\parent$.

The region where momentum $p_\unobserved$ is collinear with momentum
$p_\parent$ is defined by,
\begin{equation}
|s_{\parent\unobserved}| < \smin.
\end{equation}
In this region we introduce the hard momentum $p_\hard$ which is the
amount of the parent momentum $p_\parent$ remaining after the emission
of the unobserved collinear momentum $p_\unobserved$ such that,
\begin{eqnarray}
p_\hard = z p_\parent,&~~~~~~~&s_{a\hard} = z s_{a\parent},\nonumber
\\ p_\unobserved = (1-z)p_\parent,&~~~~~~~&|s_{a\unobserved}| =
(1-z)s_{a\parent}.
\label{eq:z}
\end{eqnarray}
In this limit the phase space factorizes,
\begin{equation}
\frac{1}{2s_{a\parent}}dP^d(a+\parent \to \unobserved+Q)
\to dP^d_{\rm col}(\parent\to \unobserved+\hard)\times
\frac{1}{2s_{a\hard}}dP^d(a+\hard \to Q),
\end{equation}
where, taking $d=4-2\e$, we find,
\begin{equation}
dP^{4-2\e}_{\rm col}(\parent\to \unobserved+\hard) =
\frac{(4\pi)^\e}{16\pi^2\Gamma(1-\e)} z\,dz\,d|s_{\parent\unobserved}|
\left[(1-z)|s_{\parent\unobserved}|\right]^{-\e} .
\label{eq:colPS}
\end{equation}
The square bracket contains the necessary factors to regulate the
poles in the matrix elements in $(1-z)$ and $s_{\parent\unobserved}$.

Combining these results we find that in the collinear limit, the full
phase space measure of interest factorizes as follows,
\begin{equation}
dP^d(a+\parent \to \unobserved+2+ \cdots +n) = dP^d_{\rm
col}(\parent\to \unobserved+\hard)\times \frac{1}{s_{a\hard}}
dP^d(a+\hard\to 2+\cdots + n).
\end{equation}

\subsection{Behavior of matrix elements}

The matrix elements also undergo a collinear factorization when one of
the final state partons is collinear with one of the initial state
partons.  Take the case where an initial state parton $\parent$ splits
into partons $\unobserved$ and $\hard$ (which participates in the hard
scattering) as in Eq.~\ref{eq:z}; then, for each ordered amplitude,
\begin{equation}
\Big | A(\ldots,\parent,\unobserved,n,\ldots) \Big |^2 \to
\hat c_F^{\parent\to \unobserved\hard}
\Big | A(\ldots,\hard,n,\ldots)   \Big |^2,
\label{eq:colM}
\end{equation}
where,
\begin{equation}
\hat c_F^{\parent\to \unobserved\hard} = \left(\frac{g^2N}{2}\right)
                     \frac{1}{|s_{\parent\unobserved}|}
\frac{\hat P_{\hard\unobserved \to \parent}(z)}{z}.
\label{eq:split}
\end{equation}
Note that the quantum numbers of the unobserved parton $u$ are determined
by the quantum numbers of the parent parton $\parent$ and the
hard-process parton $\hard$.

This is very similar to the factorization that occurs when two final
state particles are collinear. In this case, when parton $a$ (which
participates in the hard scattering) splits into a final state
collinear parton pair $1$ and $2$ then,
\begin{equation}
\Big | A(\ldots,1,2,\ldots) \Big |^2 \to \hat c_F^{12\to a}
\Big | A(\ldots,a,\ldots)   \Big |^2,
\end{equation}
where,
\begin{equation}
\hat c_F^{12\to a} = \left(\frac{g^2N}{2}\right)
                     \frac{1}{s_{12}} \hat P_{12\to a}(z),
\end{equation}
and,
\begin{equation}
p_1 = z p_a,\hskip 1cm p_2 = (1-z) p_a.
\end{equation}
The different averaging factors for initial- and final-state quarks
and gluons have been taken into account, however, we do not sum here
over different flavors of quarks participating in the hard process.

As before, the splitting functions may be either in the conventional
scheme (all particles in $d$-dimensions) \cite{AP} or in the 't Hooft
Veltman scheme (only unobserved particles in $d$-dimensions)
\cite{HV}.  In the conventional scheme, the splitting functions are
given by,
\begin{eqnarray}
\hat P_{gg\to g}(z) &=&
P_{gg\to g}(z) =
4\left(\frac{z}{1-z}+\frac{1-z}{z}+z(1-z)\right),\nonumber \\
\hat P_{qg\to q}(z) &=& \left(1-\frac{1}{N^2}\right) P_{qg\to q}(z) =
2\left(1-\frac{1}{N^2}\right)\left(\frac{1+z^2-\e
(1-z)^2}{1-z}\right),
\nonumber \\
\hat P_{q\bar q\to g}(z) &=& \frac{1}{N} P_{q\bar q\to g}(z) =
\frac{2}{N} \left(\frac{z^2+(1-z)^2-\e}{1-\e}\right).
\label{eq:AP}
\end{eqnarray}
For the splitting functions in the 't Hooft Veltman scheme, see
Ref.~\cite{HV}.

One difference from pure final state singularities is that the initial
state parton is always hard --- there is always a minimum value for $z$
imposed by demanding that a hard scattering takes place.  On the other
hand, the upper bound on $z$ is still determined by the requirement
that parton $\unobserved$ is collinear but not soft.  In other words,
$s_{\unobserved n} > \smin$, where $n$ is the neighbouring hard
parton in the ordered amplitude (see eq.~\ref{eq:colM}).

For the $g \to gg$ process, there will be contributions from two
ordered amplitudes,
\begin{equation}
 \Big | A(\ldots,m,\parent,\unobserved,n,\ldots ) \Big |^2 +\Big |
A(\ldots,m,\unobserved,\parent,n,\ldots ) \Big |^2 \to
\Big | A(\ldots,m,\hard,n,\ldots )\Big |^2,
\label{eq:col}
\end{equation}
where the order of the other hard partons in the ordered amplitude is
preserved.  The upper limit on $z$ will be different in each case
since the requirement that gluon $\unobserved$ be unobserved depends
on the adjacent momenta.  Note that in the final state case, each
ordering counts equally, however, the Bose symmetry factor takes this
into account. For processes involving collinear quarks (antiquarks),
only one ordering will contribute.  Note that only ordered amplitudes
where $\parent$ and $\unobserved$ are adjacent contribute in the
collinear limit. If they are not adjacent as in Eq.~\ref{eq:col} the
collinear limit gives a contribution of the order of the parton
resolution cut $\smin$, which is therefore negligible.  This
property is very useful since it avoids overlapping divergences for a
given ordered amplitude and this makes partial fractioning to isolate
the divergences unnecessary.

\subsection{Behavior of the cross section}

In this subsection we derive exact expressions for the crossing
functions $C_a^H(x)$ or, equivalently the effective structure function
${\cal F}_a^H(x)$ as defined in Eqs.~\ref{eq:signlo}--\ref{eq:nlo}.  We
will consider the generic process of scattering of partons $a$ and
$\hard$ to form an arbitrary final state with an invariant mass
$\sqrt{Q^2}$ (e.g.  partons only, vector boson plus partons, etc.).
The leading order cross section for the production of a vector boson
plus partons is given in Eq.~\ref{eq:siglo}.  Cross sections for other
final states are given by similar formul\ae.  The next-to-leading
order cross section is defined in Eq.~\ref{eq:signlo}, or in its
expanded form in Eq.~\ref{eq:nlo}.

The first step in the derivation of the crossing function is to
consider the initial-state collinear radiation contribution to the
next-to-leading order cross section. Consider the splitting of a
parent parton $\parent$ to a (unobserved) collinear parton
$\unobserved$, and a parton $\hard$ participating in the hard
scattering: $\parent\to \unobserved\hard $, where the invariant mass
$|s_{\parent\unobserved}| <
\smin$ so that this configuration is indistinguishable from the leading order
configuration where parton $\hard$ comes directly from the hadron.
This contributes to the next-to-leading order cross section and using
Eqs.~\ref{eq:colPS} and \ref{eq:colM} we find
\begin{equation}
d\,\sigma_{\rm initial} = \sum_{a\hard\parent} f_a^{H_1}(x_1)\left\{
f_\parent^{H_2}(y)
\hat{c}_F^{\parent\to \unobserved\hard}
dP^d_{\rm col}(\parent \to \unobserved+\hard) \delta(x_2 - z y)dy
\right\} d\sigma_{a\hard}^{LO}(x_1,x_2) dx_1 dx_2 ,
\label{eq:siginit}
\end{equation}
where, by definition, the momentum fraction $x_2$ carried by parton
$\hard$ is given by the momentum fraction $y$ of the original parton
$\parent$ multiplied by the energy fraction remaining after radiating
the unobserved parton $\unobserved$.  There is an implicit integration
over $z$ contained in the collinear phase space factor, see
Eq.~\ref{eq:colPS}.  Comparing Eq.~\ref{eq:siginit} with
Eq.~\ref{eq:nlo} gives the contribution of the initial state radiation
to the crossing function,
\begin{equation}
\alpha_s C_{\hard,\ \rm initial}^{H_2}(x_2) =
 \sum_\parent f_\parent^{H_2}(y) \hat{c}_F^{\parent\to
\unobserved\hard } \dPhase^{d}_{\rm col}(\parent \to \unobserved+\hard
) \delta(x_2 - z y) dy.
\end{equation}
Using Eq.~\ref{eq:split} and the collinear phase space factor of
Eq.~\ref{eq:colPS} gives,
\begin{eqnarray}
C_{\hard,\ \rm initial}^{H_2}(x_2) &=& -\left(\frac{N}{2\pi}\right)
\frac{1}{\Gamma(1-\e)}\left(\frac{4\pi\mu^2}{\smin}\right)^\e
\frac{1}{\e} \sum_\parent \frac{1}{4}\int^{1-z_2}_{x_2} \frac{dz}{z}
(1-z)^{-\e}\hat P_{\hard\unobserved\to \parent}(z)
f^{H_2}_\parent\left(\frac{x_2}{z}\right) \nonumber \\
&=&-\left(\frac{N}{2\pi}\right) \frac{1}{\Gamma(1-\e)}
\left(\frac{4\pi\mu^2}{\smin}\right)^\e \frac{1}{\e} \sum_\parent
\int^{1}_{x_2} \frac{dz}{z} f^H_\parent\left(\frac{x_2}{z}\right)
J_{\parent\to \hard}(z,z_2)\ ,
\label{eq:Jdefinition}
\end{eqnarray}
where we have integrated $y$ over the delta function and
$s_{\parent\unobserved}$ using the constraint
$|s_{\parent\unobserved}| < \smin$.  Here $\mu$ is an arbitrary
scale introduced to keep the strong coupling constant $\alpha_s =
g^2\mu^{-2\epsilon}/4\pi$ dimensionless in $d$-dimensions.  The upper
boundary on the $z$ integral is determined by the constraint that the
unobserved parton $\unobserved$ is not soft with respect to its
neighbouring parton $n$.  In other words, $|s_{\unobserved n}| = (1-z)
|s_{\hard n}| > \smin$.  Explicitly this gives
\begin{equation}
z < 1-\frac{\smin}{|s_{\hard n}|} = 1 - z_2.
\end{equation}
Note that $s_{\hard n}$ is only defined because of the use of the
ordered amplitudes and is different for each ordering.

Looking at the definition of the splitting functions, Eq.~\ref{eq:AP},
we see that $J_{q\to g}$ and $J_{g\to q}$ do not depend on the upper
boundary on $z$ (up to negligible corrections of $\O(\smin)$).  This
is due to the absence of a singularity in the limit that the quark or
anti-quark becomes soft.  In contrast, the $J_{g\to g}$ and $J_{q\to
q}$ functions do contain a soft singularity, arising from the limit
where the gluon becomes soft, and therefore do depend on $z_2$.  In
order to write these contributions to the crossing
functions in the second form in Eq.~\ref{eq:Jdefinition}, we use the
$(~)_+$ prescription defined by,
\begin{equation}
\left(F(z)\right)_+ = \lim_{\beta \to 0} \left(
\theta(1-z-\beta) F(z)-\delta(1-z-\beta)\int^{1-\beta}_0 F(y) \, dy \right),
\end{equation}
such that,
\begin{equation}
\int^{1-z_2}_x dz \frac{g(z)}{(1-z)^{1+\epsilon}} =
\int^{1}_x dz \frac{g(z)}{[(1-z)^{1+\epsilon}]_+}
+\left(\frac{z_2^{-\e}-1}{\e}\right)g(1),
\end{equation}
and,
\begin{equation}
\int^{1}_x dz \frac{g(z)}{[(1-z)^{1+\epsilon}]_+} =
\int^1_x dz  \frac{g(z)}{(1-z)_+} -
\e\int^1_x dz g(z)\left(\frac{\log(1-z)}{1-z}\right)_+
+{\cal O}(\e^2),
\end{equation}
\begin{equation}
\int^1_x dz  \frac{g(z)}{(1-z)_+} =
\int^1_x dz \frac{g(z)-g(1)}{1-z}+g(1)\log(1-x),
\end{equation}
\begin{equation}
\int^1_x dz g(z)\left(\frac{\log(1-z)}{1-z}\right)_+ =
\int^1_x dz \frac{g(z)-g(1)}{1-z}\log(1-z) +\frac{g(1)}{2}\log^2(1-x),
\end{equation}
provided that $g(z)$ is a function well behaved at $z=1$.

The functions $J_{\parent\to \hard}$ are thus given by,
\begin{eqnarray}
J_{g\to g}(z,z_2) &=&
\left(\frac{z_2^{-\e}-1}{\e}\right)\delta(1-z)
+ \frac{z}{[(1-z)^{1+\epsilon}]_+} + \frac{(1-z)^{1-\epsilon}}{z} +
z(1-z)^{1-\epsilon} +\O(\smin),
\nonumber \\
J_{q\to q}(z,z_2) &=& \left(1-\frac{1}{N^2}\right)
\left\{
\left(\frac{z_2^{-\e}-1}{\e}\right)
\delta(1-z)
+ \frac{1}{2}\left(\frac{1+z^2}{[(1-z)^{1+\epsilon}]_+} - \epsilon
(1-z)^{1-\epsilon} \right) \right\} +\O(\smin),
\nonumber \\
J_{q\to g}(z,z_2) &=& \frac{1}{4} \hat{P}_{g q\to q}(z)(1-z)^{-\e}
+\O(\smin), \nonumber \\ J_{g\to q}(z,z_2) &=& \frac{1}{4}
\hat{P}_{q\bar q\to g}(z)(1-z)^{-\e} +\O(\smin).
\label{eq:J}
\end{eqnarray}

The next step is to correct for the fact that we have crossed a final
state collinear cluster to the initial state. As explained in
Sec.~\ref{sec:org} this is done by subtracting the collinear factor
resulting from the splitting $\hard \to \unobserved \parent$
integrated over the final state collinear phase space (see \cite{GG}
for a detailed derivation). The contribution to the next-to leading
order cross section is given by,
\begin{equation}
d\,\sigma_{\rm final} = \sum_{a\hard\parent}
f_a^{H_1}(x_1)\left\{f_\hard^{H_2}(x_2)
\hat{c}_F^{\parent\unobserved\to \hard} dP^d_{\rm col, final}(\hard
\to \unobserved + \parent)\right\} d\sigma_{a\hard}^{LO}(x_1,x_2) dx_1
dx_2,
\end{equation}
giving,
\begin{eqnarray}
\alpha_s C_{\hard, \rm final}^{H_2}(x_2) &=& f_\hard^{H_2}(x_2)
\sum_\parent
 \hat{c}_F^{\parent\unobserved\to \hard}
\dPhase^{d}_{\rm col, final}(\hard \to \unobserved+\parent) \nonumber \\
&=& -\left(\frac{\alpha_s N}{2\pi}\right) \frac{1}{\Gamma(1-\e)}
\left(\frac{4\pi\mu^2}{\smin}\right)^\e f_\hard^{H_2}(x_2)
\frac{1}{\e}\sum_\parent I_{\parent\unobserved \to \hard}(z_1,z_2).
\end{eqnarray}
Note that the parton density function is associated with parton
$\hard$ rather than with parton $\parent$.  The integration boundaries
of the $z$ integral are again defined through the requirement that the
hard partons are resolved.  For each ordered amplitude $z_1$ and $z_2$
are given by demanding that the invariant mass of parton $\unobserved$
with both its neighbors in the particular ordering is larger than the
parton resolution cut $\smin$ so that $\unobserved$ is not soft.  In
the conventional scheme, the final state integrals over the splitting
functions, $I_{\parent\unobserved\to \hard}$, are given by,
\begin{equation}
I_{\parent\unobserved\to \hard}(z_1,z_2)=
\frac{1}{4} \int^{1-z_2}_{z_1} dz \left[ z(1-z)\right]^{-\e}
\hat P_{\parent\unobserved\to \hard}(z)
\end{equation}
where,
\begin{eqnarray}
I_{gg\to g}(z_1,z_2) & = &
\left(\frac{z_1^{-\e}+z_2^{-\e}-2}{\e}\right) -\frac{11}{6}
+\left(\frac{\pi^2}{3}-\frac{67}{18}\right)\e+\O(\e^2),\nonumber \\
I_{qg\to q}(z_1,z_2) & = & \left(1-\frac{1}{N^2}\right)
\left[ \left(\frac{z_2^{-\e}-1}{\e}\right)-\frac{3}{4}+\left(\frac{\pi^2}{6}
-\frac{7}{4}\right)\e \right] +\O(\e^2),\nonumber \\ I_{q\bar q\to
g}(z_1,z_2) & = &\frac{1}{N}\left[ \frac{1}{3}+\frac{5\e}{9} \right] +
\O(\e^2).
\label{eq:I}
\end{eqnarray}

We can now define the crossing function as a convolution integral
involving the parton density function and a crossing kernel
$X_{\parent\to \hard}(z)$ which is obtained by subtracting the final
state contribution given by Eq.~\ref{eq:I} from the initial state
contributions of Eq.~\ref{eq:J}
\begin{eqnarray}
C_\hard^H(x) &=& C^H_{\hard,\ \rm initial}(x) - C^H_{\hard,\ \rm
final}(x) \nonumber \\ &=& \sum_\parent \int^{1}_x \frac{dz}{z}
f^H_\parent\left(\frac{x}{z}\right) X_{\parent\to \hard}(z) ,
\end{eqnarray}
where the crossing kernel for specific processes is given by
\begin{eqnarray}
X_{g\to g}(z) &=& -\left(\frac{N}{2\pi}\right)\frac{1}{\Gamma(1-\e)}
\left(\frac{4\pi\mu^2}{\smin}\right)^\e \frac{1}{\e}\nonumber \\
&\times& \left( J_{g\to g}(z,z_1)+J_{g\to g}(z,z_2) - \left[ I_{gg\to
g}(z_1,z_2)+\frac{n_f}{N}I_{q\bar q\to g}(0,0)\right]
\delta(1-z)\right)\nonumber \\ &=&
-\left(\frac{N}{2\pi}\right)\frac{1}{\Gamma(1-\e)}
\left(\frac{4\pi\mu^2}{\smin}\right)^\e \nonumber \\ & \times
&\frac{1}{\e} \biggl[2\left( \frac{z}{[(1-z)^{1+\epsilon}]_+} +
\frac{(1-z)^{1-\epsilon}}{z} + z(1-z)^{1-\epsilon} \right)\nonumber \\
&& +\left( \left(\frac{11N-2n_f}{6N}\right) - \e \left(\frac{\pi^2}{3}
-\frac{67}{18} + \frac{5 n_f}{9N}\right)\right)
\delta(1-z)\biggr],\nonumber \\ X_{q\to q}(z) &=&
-\left(\frac{N}{2\pi}\right) \frac{1}{\Gamma(1-\e)}
\left(\frac{4\pi\mu^2}{\smin}\right)^\e \frac{1}{\e} \biggl( J_{q\to
q}(z,z_2) - I_{qg\to q}(0,z_2)\delta(1-z)\biggr) \nonumber \\ &=&
-\left(\frac{N}{2\pi}\right) \frac{1}{\Gamma(1-\e)}
\left(\frac{4\pi\mu^2}{\smin}\right)^\e\left(1-\frac{1}{N^2}\right)
\nonumber \\ & \times & \frac{1}{\e} \left[\
\frac{1}{2}\left(\frac{1+z^2}{[(1-z)^{1+\epsilon}]_+} - \epsilon
(1-z)^{1-\epsilon} \right) +\left(\frac{3}{4} -\e
\left(\frac{\pi^2}{6}-\frac{7}{4} \right)\right) \delta(1-z)\right]
{}. \nonumber
\label{eq:X}
\end{eqnarray}
We see that the dependence on the boundaries exactly cancels, making
the crossing function independent of the hard process.  The other two
functions, $X_{g\to q}(z)$ and $X_{q\to g}(z)$, do not receive
contributions from the final state crossing,
\begin{eqnarray}
X_{g\to q}(z) &=& -\left(\frac{N}{2\pi}\right)\frac{1}{\Gamma(1-\e)}
\left(\frac{4\pi\mu^2}{\smin}\right)^\e \frac{1}{\e} J_{g\to
q}(z,0)\nonumber \\ &=&
-\left(\frac{N}{2\pi}\right)\frac{1}{\Gamma(1-\e)}
\left(\frac{4\pi\mu^2}{\smin}\right)^\e \frac{1}{\e}
\left(\frac{1}{4} \hat P_{q\bar q\to g}(z) (1-z)^{-\e}\right),
\nonumber \\ X_{q\to g}(z) &=& -\left(\frac{N}{2\pi}\right)
\frac{1}{\Gamma(1-\e)}
\left(\frac{4\pi\mu^2}{\smin}\right)^\e\frac{1}{\e}
J_{q\to g}(z,0) \nonumber \\ &=&
-\left(\frac{N}{2\pi}\right) \frac{1}{\Gamma(1-\e)}
\left(\frac{4\pi\mu^2}{\smin}\right)^\e \frac{1}{\e}
\left(\frac{1}{4} \hat P_{gq\to q}(z) (1-z)^{-\e}\right).
\label{eq:XX}
\end{eqnarray}
Here, we have replaced endpoints that do not contribute with a zero
(that is, we have simply dropped contributions of $\O(\smin)$).

We have now derived the process independent crossing function.  They
still contain the mass singularity which has to be removed by the mass
factorization prescription. In fact in the language we have developed
here the mass factorization is done very easily as is shown in the
next subsection.

\subsection{Mass factorization}

The only physical, and therefore finite, quantity associated with
resolved partons is the effective structure function ${\cal
F}_\hard^H(x)$ as defined in Eq.~\ref{eq:F}. Conventionally, the
parton density function is made finite by renormalizing the parton
density function at the factorization scale $\mu_F$,
\begin{equation}
f_\hard^H(x) = f_\hard^H(x,\mu_F) -
\alpha_s\sum_\parent\int_x^1\frac{d\,z}{z}
f_\parent^H\left(\frac{x}{z},\mu_F\right)
R_{\parent\to\hard}(z,\mu_F) 	 + {\cal O}(\alpha_s^2)\ .
\end{equation}
This is very similar to coupling constant renormalization.  The
$\O(\alpha_s)$ term is subsequently absorbed in the crossing function
\begin{eqnarray}
{\cal F}_\hard^H(x) &=& f_\hard^H(x) + \alpha_s C_\hard^H(x)
\nonumber \\ &=& f_\hard^H(x,\mu_F) 		+ \alpha_s
C_\hard^H(x,\mu_F) 		 + {\cal O}(\alpha_s^2) ,
\end{eqnarray}
with
\begin{equation}
C^H_\hard(x,\mu_F) = \sum_\parent \int^1_x \frac{dz}{z}
f^H_\hard\left(\frac{x}{z},\mu_F\right) \biggl( X_{\parent\to
\hard}(z) 		 + R_{\parent\to \hard}(z,\mu_F)\biggr)\ ,
\end{equation}
where the mass factorization counter function absorbs the divergences
in the crossing functions. Note that the effective structure function,
${\cal F}_\hard^H$, is left unchanged by the mass
factorization and is in fact independent of the factorisation scale.
However for a cross section calculated at fixed order in perturbation
theory, we have to expand the effective structure functions explicitly
and neglect terms of ${\cal O}(\alpha_s^2)$ as was done in
Eq.~\ref{eq:nlo}. This makes the fixed order cross section
factorization scale dependent since we have to neglect the term
$\alpha_s^2(\mu_F) C_a^{H_1}(x_1,\mu_F)C_b^{H_2}(x_2,\mu_F)$.

The mass factorization counter functions $R_{\parent\to \hard}$ at the
factorization scale $\mu_F$ are given by,
\begin{eqnarray}
R^{\rm scheme}_{g\to g}(z,\mu_F) &=&
\left(\frac{N}{2\pi}\right)\left(\frac{4\pi\mu^2}{\mu_F^2}\right)^\e
\frac{1}{\Gamma(1-\e)}\frac{1}{\e} \nonumber \\ &&\times
\left\{\frac{(11N-2n_f)}{6N}\delta(1-z)
 + 2\left( \frac{z}{(1-z)_+} + \frac{(1-z)}{z} + z(1-z) \right) + \e
f^{\rm scheme}_{g\to g}(z)\right\},\nonumber \\ R^{\rm scheme}_{q\to
q}(z,\mu_F) &=&
\left(\frac{N}{2\pi}\right)
\left(\frac{4\pi\mu^2}{\mu_F^2}\right)^\e
\frac{1}{\Gamma(1-\e)}\frac{1}{\e}\left(1-\frac{1}{N^2}\right)\nonumber \\
&&\times
\left\{
\frac{3}{4}\delta(1-z)
+ \frac{1}{2}\left(\frac{1+z^2}{(1-z)_+}\right) +\e f^{\rm
scheme}_{q\to q}(z)\right\}, \nonumber \\ R^{\rm scheme}_{g\to
q}(z,\mu_F) &=&
\left(\frac{N}{2\pi}\right)
\left(\frac{4\pi\mu^2}{\mu_F^2}\right)^\e
\frac{1}{\Gamma(1-\e)}\frac{1}{\e}\left\{ \frac{1}{4}\hat P^4_{q\bar q\to g}(z)
+\e f^{\rm scheme}_{g\to q}(z)\right\}, \nonumber \\ R^{\rm scheme}_{q
\to g}(z,\mu_F) &=&
\left(\frac{N}{2\pi}\right)
\left(\frac{4\pi\mu^2}{\mu_F^2}\right)^\e
\frac{1}{\Gamma(1-\e)}\frac{1}{\e}\left\{
\frac{1}{4}\hat P^4_{g q\to q}(z)+ \e f^{\rm scheme}_{q\to g}(z)\right\},
\label{eq:counter}
\end{eqnarray}
where the four dimensional part of the splitting function is given by
$\hat P^4_{ab\to c}(z)$ and $\hat P^\e_{ab\to c}(z)$ is the $d-4$
part.  The function $f^{\rm scheme}_{\parent\to\hard}(z)$ is the
scheme dependent mass factorization term chosen such that
$f^{\MSbar}_{\parent\to \hard}(z)=0$.
The strong coupling constant in (3.32) is evaluated at the scale $\mu$,
which through coupling constant renormalisation is identified with the
renormalisation scale.   Other choices of the scale are possible, however,
provided $\alpha_s\log(\mu_R^2/\mu_F^2) \ll 1$,
the difference is of $\O(\alpha_s^2)$ and can be ignored.
Indeed, this condition is necessary in order to prevent the
appearance of large logarithms. In practice, $\mu_R$ and $\mu_F$ will
usually be chosen equal, but if they are not, the ratio $\mu_R/\mu_F$
should be small.

Combining the unrenormalized crossing functions of
Eqs.~\ref{eq:X}-\ref{eq:XX} with the counter functions of
Eq.~\ref{eq:counter} gives us the finite, renormalized crossing
functions in the $\MSbar$ scheme,
\begin{equation}
C^{H,\MSbar}_{\hard}(x,\mu_F) = \left(\frac{N}{2\pi}\right)
\left[ A^H_{\hard}(x,\mu_F)\log\left(\frac{\smin}{\mu^2_F}\right)
+ B^{H,\MSbar}_{\hard}(x,\mu_F)\right],
\label{eq:CMSbar}
\end{equation}
where the arbitrary scale $\mu$ has canceled and,
\begin{eqnarray}
A^H_{\hard}(x,\mu_F) &=& \sum_\parent A^H_{\parent\to
\hard}(x,\mu_F),\nonumber \\ B^{H,\MSbar}_{\hard}(x,\mu_F) &=&
\sum_\parent B^{H,\MSbar}_{\parent\to \hard}(x,\mu_F).
\end{eqnarray}
The finite scheme independent functions $A^H_{\parent \to
\hard}(x,\mu_F)$ are given by,
\begin{eqnarray}
A^H_{g\to g}(x,\mu_F) &=&
\int^1_x  \frac{dz}{z} f^H_g\left(\frac{x}{z},\mu_F\right)
\left\{\frac{(11N-2n_f)}{6N} \delta(1-z)
 + 2\left( \frac{z}{(1-z)_+}+ \frac{(1-z)}{z}+ z(1-z) \right)
\right\}, \nonumber \\ A^H_{q\to q}(x,\mu_F) &=&
\int^1_x  \frac{dz}{z} f^H_q\left(\frac{x}{z},\mu_F\right)
\left(1-\frac{1}{N^2}\right)
\left\{\frac{3}{4}\delta(1-z)
+ \frac{1}{2}\left(\frac{1+z^2}{(1-z)_+}\right)
\right\},\nonumber \\
A^H_{g\to q}(x,\mu_F) &=&
\int^1_x  \frac{dz}{z} f^H_g\left(\frac{x}{z},\mu_F\right)
\frac{1}{4}\hat P^4_{q\bar q\to g}(z),\nonumber \\
A^H_{q\to g}(x,\mu_F) &=&
\int^1_x  \frac{dz}{z} f^H_q\left(\frac{x}{z},\mu_F\right)
\frac{1}{4}\hat P^4_{gq\to q}(z),
\label{eq:A}
\end{eqnarray}
and the scheme dependent functions $B^{H,\MSbar}_{\parent\to
\hard}(x)$ by,
\begin{eqnarray}
B^{H,\MSbar}_{g\to g}(x,\mu_F) &=&
\int^1_x  \frac{dz}{z} f^H_g\left(\frac{x}{z},\mu_F\right)\nonumber \\
&\times &
\Biggl\{\left(\frac{\pi^2}{3}-\frac{67}{18}+\frac{5n_f}{9N}\right) \delta(1-z)
\nonumber \\
&& + 2 z\left(\frac{\log(1-z)}{(1-z) }\right)_+ +
2\left(\frac{(1-z)}{z}+ z(1-z) \right)\log(1-z)
\Biggr\}, \nonumber \\
B^{H,\MSbar}_{q\to q}(x,\mu_F) &=&
\int^1_x  \frac{dz}{z}
f^H_q\left(\frac{x}{z},\mu_F\right)\left(1-\frac{1}{N^2}\right)
\nonumber \\
&\times &
\left\{
\left(\frac{\pi^2}{6}-\frac{7}{4}\right) \delta(1-z)
-\frac{1}{2} (1-z) +
\frac{1}{2}(1+z^2)\left(\frac{\log(1-z)}{(1-z)}\right)_+ \right\},
\nonumber \\ B^{H,\MSbar}_{g\to q}(x,\mu_F) &=&
\frac{1}{4}\int^1_x  \frac{dz}{z} f^H_g\left(\frac{x}{z},\mu_F\right)
\left\{\hat P^4_{q\bar q\to g}(z)\log(1-z)
      -\hat P^\epsilon_{q\bar q\to g}(z)\right\},
\nonumber \\
B^{H,\MSbar}_{q\to g}(x,\mu_F) &=&
\frac{1}{4}\int^1_x  \frac{dz}{z} f^H_q\left(\frac{x}{z},\mu_F\right)
\left\{ \hat P^4_{gq\to q}(z)\log(1-z)
       -\hat P^\epsilon_{gq\to q}(z)\right\} .
\label{eq:B}
\end{eqnarray}
With these formulae we can calculate the crossing functions for each
set of given parton density functions. These crossing functions are
independent of the hard process. It is now straightforward to
evaluate $A_\hard(x,\mu_F)$, $B_\hard(x,\mu_F)$ and
$C_\hard(x,\mu_F)$ numerically for a given set
of input parton density functions
in a particular scheme.  We use the $\MSbar$ scheme, and in
order to give some idea of the size and relative importance of the
crossing functions we use set B1 of Ref.~\cite{MT} as input proton
$\MSbar$ parton density functions.  Furthermore, we focus on
the crossing functions associated with valence up quarks and gluons.
The distributions for down valence quarks show a similar behaviour to
the up valence quarks while the sea quarks are related to the gluonic
distributions.

First, we show the $x$ dependence of the crossing functions for
valence up quarks at a fixed scale $\mu_F = 25$~GeV in Fig.~1.  In
order to illustrate the $\smin$ dependence of $C_u$, three values of
$\smin$ have been chosen, $\smin = 1,~10$ and 100 GeV$^2$.  The
first two values are typical of the $\smin$ chosen in practical
applications (see Sec.~\ref{sec:app}) and are values where the
systematic uncertainty in evaluating the cross section is of the same
order as the uncertainty introduced by approximating the matrix
elements at small $\smin$.  Although $C_u$ does explicitly depend on
this unphysical parameter, this dependence is balanced by a growth of
the bremstrahlung contribution to the next-to-leading cross section.
Once this cancellation takes place, $\smin$ is replaced by a scale
of order of the experimental cuts.  For jets with $E_T \geq E_{Tmin}
\sim \O(15$~GeV) and a jet-jet separation of $\Delta R \sim 0.7$, this
scale is of order $E_{Tmin}^2\Delta R^2 \sim \O(100$~GeV$^2)$.  The
curves with $\smin = 100$~GeV$^2$ are representative of such a scale
and give some indication of the contribution to the physical cross
section.  As shown in Fig.~1(a), both $A_u$ and $C_u$ are negative for
some values of $x$.  Although this seems somewhat strange, $A_u$ and
$C_u$ are not directly interpretable as physical distribution; only
the complete next-to-leading order cross section as defined in
Eq.~\ref{eq:nlo} is expected to be positive (so long as higher-order
corrections are not too large).  It is worth noting that
although $A_u$ and $B^{\MSbar}_u$ are roughly similar in size,
the contribution of $A_u$ to $C_u$ is enhanced by
$\log(\smin/\mu_F^2)$ which can be large.  Therefore, particularly
for small $\smin$, the shape of $C_u$ is dictated by the scheme
independent function $A_u$, while $B_u$ is only important as $\smin
\to \mu_F^2$.  As mentioned earlier, a more physical quantity is the
effective structure function $\F_u$ defined in Eq.~\ref{eq:F}.  This
is shown in Fig.~1(b) for the same three values of $\smin$ as is the
ordinary parton density function $f_u$. We see that at large $x$,
$\F_u$ is enhanced relative to $f_u$, while at small $x$ there is a
depletion.  Furthermore, as $\smin \to \mu_F^2$, $\F_u$ approaches
$f_u$.  Note that for very small factorization scales such that
$\mu_F^2 < \smin$, then $\F_u$ is depleted at large $x$ and enhanced
at small $x$.

Fig.~2 shows the $x$ distribution for the gluonic crossing functions.
These crossing functions receive contributions from both $g \to gg$
and $q\to gq$ splitting functions, and, due to the soft gluon poles
that are present, $A_g$ and $B_g$ both grow at small $x$.  As a
consequence, $C_g$ is negative in this region.  However, for large
$\smin$, there is a significant cancellation between $A_g$ and $B_g$
so that $C_g$ is less singular.  This is reflected in Fig.~2(b) where
$\F_g$ and $f_g$ are shown for the gluon.  At small $x$ there is a
dramatic softening of the growth of the gluon density function.  This
depletion is entirely consistent with the depletion of the up valence
distribution discussed above.  Similarly, at large $x$, there is a
small enhancement.

It is also instructive to study the scale dependence of the crossing
functions.  This dependence is present in the input parton density
functions and hence $A$ and $B$ and through the
$\log(\smin/\mu_F^2)$ factor multiplying $A$ in Eq.~\ref{eq:CMSbar}.
$\F$ contains an additional $\mu_F$ dependence from the strong
coupling constant evaluated at the factorisation scale which we take
to be equal to the renormalisation scale.  Fig.~3 shows the $\mu_F$
dependence for the up valence density functions at $x=0.05$.  $A_u$ is
a slowly decreasing function, much smaller than $B_u$ while both $A_u$
and $B_u$ are positive for $\mu_F < 1000$~GeV.  As a consequence, for
$\smin < \mu_F^2$, $C_u$ is relatively small due to a cancellation
between the two terms.  This is not the case for smaller scales where
$C_u$ can be quite large.  Fig.~3(b) again shows the ordinary parton
density function $f_u$ with the effective structure function $\F_u$.
For large $\mu_F$, where $C_u$ is small, $f_u$ and $\F_u$ are very
similar in size, while at small scales we can see very large
differences.

The scale dependence of the gluonic crossing functions is shown in
Fig.~4.  Unlike the up valence quark case, $A_g$ becomes negative at a
scale of $\mu_F = 7$~GeV.  As a consequence, $A_g$ and the rather
large $B_g$ combine together coherently to form $C_g$ which grows
logarithmically at large scales.  Furthermore, the effective structure
function $\F_g$ is always significantly larger than the ordinary
parton density function $f_g$.

\newpage
\figure{The valence quark density functions (a) $A_u$, $B_u$ and $C_u$ and
(b) $\F_u$ and $f_u$ as a function of $x$ for $\mu_F = 25$~GeV.  $C_u$
and $\F_u$ are shown for $\smin = 1,~10$ and 100~GeV$^2$.\\
\vspace{24cm}
\includegraphics{x.upvabc.25.ps}
\includegraphics{x.upvf.25.ps} }
\newpage
\figure{The gluon density functions (a) $A_g$, $B_g$ and $C_g$ and
(b) $\F_g$ and $f_g$ as a function of $x$ for $\mu_F = 25$~GeV.  $C_g$
and $\F_g$ are shown for $\smin = 1,~10$ and 100~GeV$^2$. \\
\vspace{24cm}
\includegraphics{x.gluabc.25.ps}
\includegraphics{x.gluf.25.ps} }
\newpage
\figure{The valence quark density functions (a) $A_u$, $B_u$ and $C_u$ and
(b) $\F_u$ and $f_u$ as a function of the factorisation scale $\mu_F$
in GeV for $x=0.05$.  $C_u$ and $\F_u$ are shown for $\smin = 1,~10$
and 100~GeV$^2$. \\
\vspace{24cm}
\includegraphics{q.upvabc.05.ps}
\includegraphics{q.upvf.05.ps} }
\newpage
\figure{The gluon density functions (a) $A_g$, $B_g$ and $C_g$ and
(b) $\F_g$ and $f_g$ as a function of the factorisation scale $\mu_F$
in GeV for $x=0.05$.  $C_g$ and $\F_g$ are shown for $\smin = 1,~10$
and 100~GeV$^2$. \\
\vspace{24cm}
\includegraphics{q.gluabc.05.ps}
\includegraphics{q.gluf.05.ps} }
\section{Applications}
\label{sec:app}

We turn next to the construction of an example of a next-to-leading
order cross sections for jet production at hadron colliders.  Let us
focus on the process,
\begin{equation}
V \to q\bar q + n~g,
\label{eq:Vng}
\end{equation}
for which the lowest order matrix element is given by,
\begin{equation}
\M_V (Q_1;1,\ldots ,n;\overline{Q}_2|P) =
\widehat{\S}_\mu(Q_1;1,\ldots ,n;\overline{Q}_2)V^\mu.
\end{equation}
Here $V^\mu$ represents either the vector boson polarization vector, the
lepton current which created the vector boson, or the leptonic
decay products of the vector boson,
while $\widehat{\S}_\mu$ is the
hadron current. Both currents depend on the particle helicities which we
suppress throughout
and the particle momenta which we
denote by $P$ for the vector boson,  $Q_1$, $\overline{Q}_2$
for the quarks and $K_1,\ldots,K_n$ for the gluons. These momenta satisfy
the momentum conservation relation,
\begin{equation}
P^\mu = Q_1^\mu + \overline{Q}_2^\mu + K_1^\mu
+ \cdots + K_n^\mu.
\end{equation}
In addition, the hadron current depends on the colors of the gluons
$a_1,\ldots,a_n$ and the quarks $c_1$,$c_2$.

The hadron current $\widehat{\S}_\mu$ may be decomposed according to
the different allowed color structures \cite{six,MPX,cod},
\begin{equation}
\widehat{\S}_\mu(Q_1;1,\ldots ,n;\overline{Q}_2)
=ieg^n\sum_{P(1,\ldots,n)} (T^{a_1}\cdots T^{a_n})_{c_1c_2}
\S_\mu(Q_1;1,\ldots ,n;\overline{Q}_2),
\end{equation}
where $\S_\mu$ represents the colorless ordered amplitude in which the gluons
are emitted in an ordered way from the quark line.  Note that the prefactor
associated with each $\S_\mu$ is also ordered according to the color of the
gluons.  These color factors form a complete basis
 and therefore each $\S_\mu$ is gauge invariant with respect to the gluons.
Although the full hadron current $\widehat{\S}_\mu$ is invariant under
permutations of the gluons, the ordered amplitude is not.  This property
is recovered by summing the ordered amplitudes over the $n!$ gluon
permutations, $P(1,\ldots,n)$.

One advantage of using this color decomposition is that the squared matrix
elements summed over helicities and colors have a very systematic structure,
\begin{eqnarray}
|\M_V|^2 = \Big |\widehat{\S}_\mu V^\mu\Big |^2 & = &
e^2 \left(\frac{g^2N}{2}\right)^n
\left(\frac{N^2-1}{N}\right)
\left[ \sum_{P(1,\ldots,n)}\Big |\S_\mu V^\mu\Big |^2 +
\O\left(\frac{1}{N^2}\right) \right],\nonumber \\
\end{eqnarray}
where $n\geq 1$ counts the number of gluons. On the right hand side, we have
expanded in the number of colors.
The terms subleading in the number of colors are related to matrix elements
with abelian couplings.  For example, when $n=2$,
\begin{eqnarray}
\lefteqn{\Big |\widehat{\S}_\mu(Q_1;1,2;\overline{Q}_2)
 V^\mu\Big |^2  =
e^2 \left(\frac{g^2N}{2}\right)^2
\left(\frac{N^2-1}{N}\right)}\nonumber \\
&\times&\left[
\sum_{P(1,2)}
\Big |\S_\mu(Q_1;1,2;\overline{Q}_2) V^\mu\Big |^2
-\frac{1}{N^2}
\Big |\S_\mu(Q_1;\tilde{1},\tilde{2};\overline{Q}_2) V^\mu\Big |^2
\right],
\end{eqnarray}
where,
\begin{equation}
\S_\mu(Q_1;\tilde{1},\tilde{2};\overline{Q}_2)
= \S_\mu(Q_1;1,2;\overline{Q}_2) +\S_\mu(Q_1;2,1;\overline{Q}_2),
\end{equation}
and the contribution from the triple gluon vertex drops out. The Feynman
graphs contributing to $\S_\mu(Q_1;\tilde{1},\tilde{2};\overline{Q}_2)$ are
therefore
those for $V \to q\bar q + 2~\gamma$ or $V \to q\bar q + g+\gamma$.

The ordered amplitudes also have special properties in the soft gluon or
collinear parton limits which allow us to isolate the singular
regions using the parton resolution parameter $\smin$ \cite{soft,GG}.
These divergences are proportional to the lowest order squared
ordered amplitudes as are the virtual divergences.
Therefore we can combine them directly and,
due to the Kinoshita-Lee-Nauenberg \cite{KLN,GS}
theorems, obtain  a finite result  after the usual coupling constant
renormalization. The finite next-to-leading order matrix elements can be
written,
\begin{eqnarray}
\Big |\widehat{\S}_\mu V^\mu\Big |^2_F
=
e^2 \left(\frac{g^2N}{2}\right)^n
\left(\frac{N^2-1}{N}\right)
\sum_{P(1,\ldots ,n)}&&
\Biggl [ \K(Q_1;1,\ldots ,n;\overline{Q}_2)
{}~\Big |\S_\mu(Q_1;1,\ldots ,n;\overline{Q}_2)V^\mu\Big |^2\nonumber \\
&&~~~+\F(Q_1;1,\ldots,n;\overline{Q}_2)
+\O\left(\frac{1}{N^2}\right)~\Biggr ],
\end{eqnarray}
where $\F(Q_1;1,\ldots,n;\overline{Q}_2)$ is the finite contribution from
the virtual graphs. The dynamical ordered $\K$ factor is given by,
\begin{eqnarray}
\K(Q_1;1,\ldots,n;\overline{Q}_2)
& = & \left(\frac{\alpha_s(\mu_R^2)N}{2\pi}\right)
\biggl[ \sum_{ij}\left\{-\log^2\left(\frac{|s_{ij}|}{\smin}\right)
+\frac{\pi^2}{2}\left(\Theta(s_{ij})-\frac{2}{3}\right) \right\}\nonumber \\
&&~~~~+\frac{3}{4}\log\left(\frac{|s_{Q_11}|}{\smin}\right)
+\frac{3}{4}\log\left(\frac{|s_{n\overline{Q}_2}|}{\smin}\right)
+\frac{67n-9}{18}-\frac{5nn_f}{9N}\biggr]\nonumber \\
&+&\alpha_s(\mu_R^2)b_0n\log\left(\frac{\mu_R^2}{\smin}\right)
+ \O(\e) + \O(\smin),
\label{eq:Kn}
\end{eqnarray}
where the sum runs over all $(n+1)$ color-connected pairs, that is,
$ij = Q_11, 12,\ldots ,n\overline{Q}_2$, and where $b_0$ is the
one-loop coefficient of the QCD beta function.  Note that $\K(Q_1;1,\ldots
,n;\overline{Q}_2)$ depends explicitly on the parton resolution
parameter $\smin$. The renormalisation scale $\mu_R$ is the scale at
which the $\MSbar$ counter term is subtracted.  For vector boson decay
all $s_{ij} > 0$ so that $\Theta(s_{ij})=1$.  However, when we cross
partons into the initial state this will no longer be true and it is
necessary to maintain the explicit analytic continuations of the
$\log^2$ terms.  Up to this point, we have continued
 both matrix elements and phase space into $d=4-2\e$
 using dimensional regularization.
In Eq.~\ref{eq:Kn}, we now see that all singularities have
cancelled explicitly and we may therefore take the 4-dimensional
limit.  This means that the squared ordered amplitudes, $\Big | \S_\mu
V^\mu\Big |^2$, may be evaluated in 4-dimensions and not in
$d$-dimensions, thus simplifying the calculation dramatically.
Similarly, it is not necessary to extend the jet algorithm to
$d$-dimensions as in the work of Ellis, Kunszt, and Soper \cite{EKS}.

Keeping all orders in the number of colors presents no problems.  For example,
the effective matrix elements for $V \to q\bar q + 2~g$ at next-to-leading
order are given by,
\begin{eqnarray}
\lefteqn{\Big |\widehat{\S}_\mu V^\mu\Big |^2_F =
 e^2\left( \frac{g^2N}{2} \right)^2 \left( \frac{N^2-1}{N} \right)}\nonumber \\
&\times&\Biggl[
\sum_{P(1,2)}
\left( \K(Q_1;1,2;\overline{Q}_2)-\frac{1}{N^2}\K(Q_1;\overline{Q}_2)\right)
\Big |\S_\mu(Q_1;1,2;\overline{Q}_2) V^\mu\Big |^2\nonumber \\
& &
-\frac{1}{N^2}
\left( \K(Q_1;1 ;\overline{Q}_2)+\K(Q_1;2;\overline{Q}_2)
-\left(1+ \frac{1}{N^2}\right)\K(Q_1;\overline{Q}_2) \right)
\Big |\S_\mu(Q_1;\tilde{1},\tilde{2};\overline{Q}_2) V^\mu\Big |^2
\Biggr],\nonumber \\
& &  +\F(Q_1;1,2;\overline{Q}_2),
\end{eqnarray}
where $\K(Q_1;\overline{Q}_2)$, $\K(Q_1;1 ;\overline{Q}_2)$
and $ \K(Q_1;1,2;\overline{Q}_2)$
are given by Eq.~\ref{eq:Kn} with $n = 0, 1$ and 2 respectively.

\subsection{$ H_1 H_2 \to V + 0$ jets}

As a first application of \ref{eq:nlo} and the
next-to-leading order crossing approach, let us
consider the production of a vector boson in hadron-hadron collisions followed
by the decay of the vector boson in the absence of jets.
The relevant parton-level processes are,
\begin{equation}
q\bar q \to V,
\label{eq:v}
\end{equation}
along with the bremstrahlung processes,
\begin{equation}
q\bar q \to V + g,~~~~~qg \to V + q, ~~~~~  g\bar q \to V + \bar q,
\label{eq:v1}
\end{equation}
for which the generic cross section is given by,
\begin{equation}
d\,\sigma_{ab}(x_1,x_2) =  \frac{\Phi_{ab}}{2\,s_{ab}}
           \left|{\cal M}_{ab} \right|^2 \
                          d\,P(ab\to V\ +\ 0,~1\,\mbox{partons}).
\end{equation}
The spin and colour averaging factors, $\Phi_{ab}$ are given by,
\begin{equation}
\Phi_{q\bar q} = \frac{1}{4N^2},
{}~~~~~~~~\Phi_{qg} = \Phi_{g\bar q} = \frac{1}{4N(N^2-1)},
{}~~~~~~~~\Phi_{gg} = \frac{1}{4(N^2-1)^2}.
\end{equation}
The lowest order matrix elements for these processes are related
to those for Eq.~\ref{eq:Vng} with $n=0$ and 1 through the usual crossing
relations. In other words, the momenta and helicity of crossed particles
are reversed.  For example,
\begin{equation}
\Big | \M_{q\bar q}(Q_1;1,\ldots,n;\overline{Q}_2|P)\Big |^2 =
\Big | \M_{V}(-Q_1;1,\ldots,n;-\overline{Q}_2|-P)\Big |^2,
\label{eq:qbarq}
\end{equation}
and,
\begin{equation}
\Big | \M_{q g}(Q_1;1,\ldots,n;\overline{Q}_2|P)\Big |^2 =
- \Big | \M_{V}(Q_1;-1,\ldots,n;-\overline{Q}_2|-P)\Big |^2.
\label{eq:qg}
\end{equation}
An explicit form for these matrix elements and hence $
|\S_\mu(Q_1;\overline{Q}_2) V^\mu  |^2$
and $  |\S_\mu(Q_1;1;\overline{Q}_2) V^\mu |^2$ using spinor language
is given in Appendix A of \cite{GG}.
Full details of how crossing affects the spinors is given in
Appendix E of Ref.~\cite{crohel}.

At next-to-leading order, the finite effective matrix elements for
Eq.~\ref{eq:Vng} with $n=0$,
 are given by,
\begin{equation}
\Big |\M_V(Q_1;\overline{Q}_2|P) \Big |^2_F   =
e^2 N \left(1-\frac{1}{N^2}\right)\left[ \K(Q_1;\overline{Q}_2)
\Big |\S_\mu(Q_1;\overline{Q}_2) V^\mu\Big |^2
+\F(Q_1;\overline{Q}_2)\right],
\end{equation}
where, because of our choice for the assignment of the finite pieces
between $\K$ and $\F$,
\begin{equation}
\F(Q_1;\overline{Q}_2)=0.
\end{equation}
The dynamical $\K$ factor is given by Eq.~\ref{eq:Kn} with $n=0$.
Because of Eq.~\ref{eq:nlo}, we may cross this in exactly the same manner
as the tree level matrix elements of Eq.~\ref{eq:qbarq}.
The only subtlety is in the analytic continuation of the $\log^2$ terms
in $\K$ which we have written out explicitly in Eq.~\ref{eq:Kn}.

It is now straightforward to construct a Monte Carlo program to
evaluate the fully differential cross section numerically.  In
particular, the vector boson decays are easily included which allows
experimental cuts to be placed directly on the observed leptons.  It
is important to note that the phase space is restricted to regions
where all partons are resolved.  In other words, any pair of partons
must have an invariant mass larger than the parton resolution
parameter, $| s_{ij}| > \smin$.  What this means in practical terms is
that the bremstrahlung contribution to the cross section grows as
$\log^2(\smin)$.  This is balanced by the explicit $-\log^2(\smin)$ in
$\K$ such that the total cross section should be independent of the
unphysical $\smin$ provided (a) that $\smin$ is small enough that the
soft and collinear approximations are valid and (b) that $\smin$ is
not so small that the numerical cancellation between the two
contributions becomes unstable.
Fig.~5 shows that the $\O(\alpha_s)$ $W+0$~jet or
$\ell^\pm + E_T^{\rm missing} + 0 $ jet cross section is essentially
independent of $\smin$ over a wide range of $\smin$.
In general, one wants to choose the largest $\smin$ possible, to
minimize the running time of the program.  In this case,  a
reasonable value to choose is $\smin = 10$~GeV$^2$.  (One must be
careful to note that for certain distributions, in particular
infrared-sensitive ones, a smaller $\smin$ might be required for
some values of the relevant kinematic variables.)

One quantity of interest is the dependence of the $W+0$ jet
cross section on the choice of experimental cuts.
In principle, including higher orders mimics more accurately the
correct dependence.  At leading order, with `standard' CDF cuts,
\begin{equation}
\sigma(W+0~{\rm jets}) = 0.78^{+0.01}_{-0.03} ~{\rm nb},
\end{equation}
This does not depend on the jet defining cut, $E_{T\rm min}^{\rm jet}$,
since at leading order there is no parton in the final state.
At next-to-leading order, this is no longer true and in Fig.~6 we
show the next-to-leading order $W+0$~jet cross section as a function of
$E_{T\rm min}^{\rm jet}$ for the same range of scales.
As $E_{T\rm min}^{\rm jet}$ becomes large, this cross-section approaches the
inclusive $W$ cross section.

\subsection{$ H_1 H_2 \to V + 1$ jet}

We now turn to vector boson production in association with a single
jet.  As before, the structure of the next-to-leading cross section is
described by Eq.~\ref{eq:nlo}, however the contributing parton-level
processes include both those of Eq.~\ref{eq:v1} as well as,
\begin{equation}
q\bar q \to V + q\bar q,~~~~   q\bar q \to V + gg,
{}~~~~qg\to V + q g, ~~~~ g\bar q \to V + g \bar q, ~~~~gg\to V + q\bar q.
\label{eq:v2}
\end{equation}
As before, the lowest order matrix elements for Eq.~\ref{eq:v2} are obtained
by crossing the matrix elements given in Appendix A of \cite{GG}.

The finite next-to-leading order matrix elements are given by,
\begin{eqnarray}
\lefteqn{\Big |\M_V(Q_1;1;\overline{Q}_2|P)  \Big |^2_F  =
e^2 \left(\frac{g^2N}{2}\right)
\left(\frac{N^2-1}{N}\right)}\nonumber \\
&\times &
\left[\left(\K(Q_1;1;\overline{Q}_2)-\frac{1}{N^2}\K(Q_1;\overline{Q}_2)\right)
\Big |\S_\mu(Q_1;1;\overline{Q}_2) V^\mu\Big |^2
+\F(Q_1;1;\overline{Q}_2)\right],
\end{eqnarray}
where $\K(Q_1;1;\overline{Q}_2)$ is given by Eq.~\ref{eq:Kn}  with $n=1$.
The finite one loop contributions are given by Eqs.~A.42-A.46 of \cite{GG}.
Once again, it is trivial to cross both $\K$ and
$\Big |\S_\mu(Q_1;1;\overline{Q}_2) V^\mu\Big |^2$.  However,
some care must be taken in crossing $\F$ since although crossing
the helicity structure in Eq.~A.43 is straightforward, the coefficients
$\alpha_i$, $\beta_i$ and $\delta_i$ in
Eq.~A.44 are expressed in terms of the function $R(x,y)$ (Eq.~A.45) which
has been written
assuming that $0 \leq x,y \leq 1$.  For crossed processes this is no longer
true.
For example, when,
\begin{equation}
x < 0,~~~y< 0,
\end{equation}
in which case,
\begin{eqnarray}
R(x,y)& = &-\frac{1}{2}\log^2(1-x)-\frac{1}{2}\log^2(1-y)
  +\frac{\pi^2}{2}\nonumber \\
&&          -{\rm Li}_2\left(\frac{1}{1-x}\right)
            -{\rm Li}_2\left(\frac{1}{1-y}\right).\nonumber \\
\end{eqnarray}
Alternatively, if,
\begin{equation}
x < 0,~~~y>1,
\end{equation}
in which case,
\begin{eqnarray}
R(x,y)& = &-\frac{1}{2}\log^2(1-x)+\frac{1}{2}\log^2(y)
+\log\left(\frac{-x}{y-1}\right)\log(y)
  +\frac{\pi^2}{2}\nonumber \\
&&          -{\rm Li}_2\left(\frac{1}{1-x}\right)
            +{\rm Li}_2\left(\frac{1}{y}\right).\nonumber \\
\end{eqnarray}
We have checked that crossed finite virtual contributions
agree with the results of Gonsalves et al.\ \cite{Gon} when the
vector boson decay current is replaced by its polarisation vector.

With these crossed matrix elements, we can
construct a Monte Carlo program to numerically
evaluate the fully differential vector boson plus one jet cross section at
next-to-leading order.  As before, the vector boson
decays are easily included which allows experimental cuts to be placed directly
on the observed leptons.
To demonstrate that the $W+1$ jet cross section is essentially
independent of the
unphysical parameter $\smin$, Fig.~7 shows that the $\O(\alpha_s^2)$
$W+1$~jet cross section as a function of $\smin$.
For $\smin$ in the range 1-10~GeV$^2$, the cross section is
not dominated by systematic errors and does not depend on $\smin$.
We therefore set $\smin = 10$~GeV$^2$ as in the $W+0$~jet case.

An important issue in $W$ + jets events is the signficance of corrections to
leading-order results.  For most quantities, radiative effects should
be small so that we can rely on leading order to describe the basic features
of the data even though
the overall normalization of the cross section is uncertain.
One of the most fundamental distributions is the jet transverse momentum
distribution which is shown in Fig.~8.

For the leading order results we choose two renormalisation scales,
$\mu_R = M_W/2$ and the total invariant mass of the event, $\mu_R =
\hat s/2$.  The first scale is the smallest scale available and
generates the hardest transverse momentum distribution while the
second scale is the hardest scale available and leads to the softest
transverse momentum spectrum.  The band defined by these two scales
represents the range of leading order predictions.  The factor $1/2$
is present in the choice of scale so that the corresponding total
cross sections at leading order (0.117 nb and 0.106 nb) are close to,
and bracket, the next-to-leading order result of 0.113 nb, which is
essentially independent of the renormalization scale (here taken to be
$M_W$).  As can be seen in Fig.~8, the next-to-leading order
distribution is somewhat softer than the leading-order
results.  In fact it is even softer than leading order with the
largest scale. This implies that the standard jet algorithm does not
take into account an important effect thereby leading to large
radiative effects for high transverse momentum jets.

{}From a physical point of view, it is clear why the spectrum softens
more dramatically than can be expected from leading order
with the standard jet algorithm.
For high transverse energy jets, the accompanying soft radiation
increases with the energy of the jet.
Therefore, with a fixed transverse momentum cut (
in this case $E_{T \rm min}^{\rm jet}=15$ GeV),
it is easier for the soft radiation in the event to fluctuate so that it passes
the minimum transverse momentum threshold and subsequently be counted as
a extra jet.
Since we are looking at the exclusive jet cross section
this event will be removed from the $W$ + 1 jet cross section and
added to the $W$ + 2 jet cross section.
This effect gets more severe when the jet transverse energy gets
larger leading to a depletion of the $W$+1 jet cross section
and  a softening of the  transverse momentum distribution.
At leading order this effect is not modelled at all because the
leading order prediction associates all the energy with the
jet and allows no soft radiation outside the jet cone.
In contrast, at next-to-leading order the
hadronic energy around the jet cone is modelled,
allowing the generation of softer jets within the tail of
high transverse energy jets.

{}From a more mathematical point of view, it is also clear what happens
in the exclusive jet cross section with high transverse momentum jets.
{}From Eq.~(4.9), we can see that the high transverse momentum jets
generate correction terms of order $-\alpha_s \log(E_{T}^{\rm
jet}/E_{T \rm min}^{\rm jet})$ which become large if the $E_{T}^{\rm
jet}$ is much larger than the minimal transverse energy.  This is
undesirable since it implies large radiative effects which are due
entirely to the jet algorithm itself.  In principle, the jet algorithm
should minimize these effects in order to be able to compare theory
with experiment.

This requires a slight modification in the jet algorithm.  By scaling
the minimal transverse momentum cut with the hardness of the event
(e.g. summed scalar energies or total invariant mass), we allow the
hard jets to radiate accompanying soft energy without generating
additional small jets. That is, we allow the jet to `vent' its energy
without producing a large number of soft jets.  Events which formerly
contained $W$ + 2 jets where one of the jets is relatively soft are
now counted as $W$ + 1 jet events, thus increasing the $W$+1 jet cross
section at high transverse momentum.  For example, by demanding $E_{T
\rm min}^{\rm jet} = \lambda \sqrt{\hat s}$, the correction term is
$-\alpha_s \log^2(\lambda)$ where we can now choose the constant
$\lambda$ and thus control the size of the corrections.

To demonstrate this effect we show in Fig.~9 the factor
which we need to multiply the leading order distribution
(with $\mu_R = \hat s/2$)
to obtain the next-to-leading order result.
For the fixed $E_{T \rm min}^{\rm jet}$, we see a sizeable correction
that depends strongly on the jet transverse energy.
However, if we take a scaling $E_{T \rm min}^{\rm jet} =
{\rm ~max}(15$~GeV, 0.1 $\sqrt{\hat s})$,
we get a result very close to
leading order with only a small enhancement for soft transverse energy
jets. For the high transverse energy jets the next-to-leading order prediction
is well described by the leading order result.

\newpage
\figure{The $\smin$ dependence of the $W+0$~jet cross section
for `standard' CDF cuts; $E_{T}^{\rm jet} \geq 15$~GeV,
$E_{T}^\ell \geq 20$~GeV, $E_T^{\rm missing} > 20$~GeV,
$|\eta^{\rm jet}| \leq 2$,
$|\eta^\ell| \leq 1$ and a jet cone size
$\Delta R = \sqrt{\Delta\phi^2+\Delta\eta^2} \leq 0.7$.
The structure functions are set B1 of \cite{MT} while the factorisation
and renormalisation scales are $\mu_F = \mu_R = M_W$.
For input parameters we choose $M_W = 80$~GeV, $\Gamma_W = 2$~GeV,
$\sin^2\theta_W = 0.23$ and $\alpha_s(M_W) = 0.1108$.\\
\vspace{15cm}
\includegraphics{ydep_0j.ps}
}
\newpage
\figure{The NLO $W+0$~jet cross section as a function of the jet defining cut
$E_{T\rm min}^{\rm jet}$ for $\mu_F=\mu_R = 2M_W,~M_W$ and $M_W/2$.
\\
\vspace{12cm}
\includegraphics{struc_1.ps}
}
\newpage
\figure{The $\smin$ dependence of the $W+1$~jet cross section
for `standard' CDF cuts.
The structure functions are set B1 of \cite{MT} while the factorisation
and renormalisation scales are $\mu_F = \mu_R = M_W$.\\
\vspace{12cm}
\includegraphics{ydep_1j.ps}
}
\newpage
\figure{The $\O(\alpha_s)$ and $\O(\alpha_s^2)$
jet and $W$ boson transverse momentum distribution
for `standard' CDF cuts.\\
\vspace{12cm}
\includegraphics{pt_log.ps}
}
\newpage
\figure{The ratio of next-to-leading order to leading order
jet transverse momentum distributions for a fixed
$E_{T\rm min}^{\rm jet} = 15$~GeV  cut  and for a scaled
$E_{T\rm min}^{\rm jet} = {\rm ~max}(15$~GeV, $0.1\times \sqrt{\hat s})$ cut
as a function of the jet transverse momentum.\\
\vspace{12cm}
\includegraphics{rat_1_1.ps}
}

\section{Conclusions}
\label{sec:con}

The main theme in this paper is the extension of the general method of
ref.~\cite{GG}, for dealing with final state collinear and infrared
divergences, to include partons in the initial state.  For final state
partons, the soft and collinear divergences from the bremstrahlung
process are isolated using a parton resolution parameter $\smin$.
These divergences are proportional to lowest order matrix elements and
can be combined directly with the divergences from the virtual graphs
to give a finite cross section, Eq.~\ref{eq:final}, which depends on a
dynamical $\K$ factor multiplying the lowest-order term, along
with a finite one loop
contribution $\F$.  In order to extend this to incorporate initial
state partons, we have extended the tree level concept of crossing to
next-to-leading order processes.  This is achieved by (a) through the
analytic continuation of the dynamical factor $\K$ (Eq.~\ref{eq:Kn})
and the finite one loop contribution $\F$ (see Sec.~\ref{sec:app})
into the physical region and (b) through the introduction of universal
crossing functions, Eq.~\ref{eq:C}, which are essentially convolutions
of the structure functions with the Altarelli-Parisi splitting
functions.  Together, $\K$, $\F$ and $C(x)$ form a set of finite
building blocks with which one can calculate next-to-leading order
cross sections.  This is summarized in Eq.~\ref{eq:nlo}; because the ordered
factorization of the soft poles and the factorization of the
collinear poles, is independent of the hard process, this equation
represents the general cross section for any hadronic process.

As an explicit example of this method, in Sec.~\ref{sec:app},
we have taken the next-to-leading order
matrix elements relevant for $e^+e^- \to 2, 3$ and 4 partons \cite{GG} and
crossed two of the partons into the initial state to obtain the cross section
for,
\begin{equation}
p\bar p \to W^\pm/Z + 0,~1~{\rm jets} \to \ell\bar\ell +  0,~1~{\rm jets},
\end{equation}
at next-to-leading order.
The phase space is evaluated numerically with the constraint
that all $|s_{ij}| > \smin$ and all final state
lepton correlations are retained.
This makes it possible to implement jet algorithms,
detector acceptance effects, and other constraints numerically,
yielding a very flexible Monte Carlo programs
as we discussed in Sec.~\ref{sec:app}.
One should verify that the cross section is independent of the
unphysical parameter $\smin$. For our Monte Carlo simulations this is
indeed the case, see Figs.~5 and~7.

It is important to note that throughout this paper we have discussed
cross sections that are exclusive in the number of jets.  As a result,
we are interested in calculating the $W+0$ jet cross section, rather than the
inclusive $W$ cross section which can only be identified with the
$W+0$ jet cross section as the mimimal transverse energy cut of the jet becomes
very large.
Similarly,  we study the $W+1$ jet cross section rather than the transverse
momentum distribution of the $W$ boson which is not directly measurable.
As shown in Fig.~8, this is not the same as the jet
transverse momentum distribution at next-to-leading order.
At large $E_{T}^{\rm jet}$, the next-to-leading order jet $p_T$
distribution is significantly softened.
This is because the existing jet algorithm  generates an artificially
high jet multiplicity  in events containing a very hard jet
by restricting the hadronic radiation around the primary jet in the
exclusive jet cross section.
By modifying the jet algorithm as described in Sec.~\ref{sec:app}, these
large radiative effects can be removed.
A detailed study of the implications of next-to-leading order corrections
to vector boson production in association with
0,~1 jets at Fermilab energies is currently in progress \cite{GGKmor,GGK2}.

The method we have presented here considerably simplifies the
structure of next-to-leading order QCD corrections to hadronic
processes.  It also makes comparison with experiment more direct
through the use of Monte Carlo simulations. Once technical problems
associated with five point loop diagrams are solved, it should be
posssible to use these methods to compute processes such as $e^+e^-
\to 4$~jets, its crossing $p\bar p \to W^\pm/Z + 2$~jets and $p\bar
p\to 3$~jets at next-to-leading order possible.  These multijet cross
sections are important for experiments at LEP and Fermilab because
event rates are high and can be studied in great detail.

\acknowledgements

WTG is happy to acknowledge financial support from the Texas National Research
Laboratory Commission.  EWNG would like to thank the Fermilab Theory group
for their kind hospitality.  We are grateful to Eric Laenen and Keith Ellis for
stimulating discussions and for providing a FORTRAN routine for the evaluation
of the strong coupling constant.  We also thank the Fermilab lattice group
for providing us with computer time on the Fermilab ACPMAPS parallel machine.
\\

\newpage


\begin{thebibliography}{99}
\bibitem{foot1}Including higher orders would give contributions where this
parton has generated a `shower'; the jet clustering will nonetheless
reconstruct closely the leading-order estimate of the jet axis and energy
(assuming an ideal detector), as a consequence of energy-momentum conservation.
\bibitem{foot2}Note that this approach is different from that
described in \cite{ERT,KS,MNR} and we refer the reader to the introduction of
\cite{KS} for a lucid explanation of the difference between the two methods.
\bibitem{foot3}Although the parton resolution parameter is
unphysical because partons are not
directly observable, this concept parallels the way in
which a  calorimeter resolves physical particles.
\end{thebibliography}
\end{document}